\documentclass[twocolumn,showpacs,preprintnumbers,amsmath,amssymb]{revtex4}
\usepackage{graphicx}
\usepackage{txfonts}
\usepackage{epsfig}
\usepackage{graphicx}
\usepackage{dcolumn}
\usepackage{bm}
\begin{document}
\title{\large Cosmological Models in the Generalized Einstein Action}
\author{Arbab I. Arbab}
\email{aiarbab@uofk.edu} \affiliation{Department of Physics,
Faculty of Science, University of Khartoum, P.O. Box 321, Khartoum
11115, Sudan,}\affiliation{Department of Physics and Applied
Mathematics, Faculty of Applied Sciences and Computer, Omdurman
Ahlia University, P.O. Box 786, Omdurman, Sudan}
\date{\today}
\begin{abstract}
We have studied the evolution of the Universe in the generalized
Einstein action of the form $R+\beta \ R^2$, where $R$ is the scalar
curvature and $\beta=\rm\ const.$. We have found exact cosmological
solutions that predict the present cosmic acceleration. These models
also allow an inflationary de-Sitter era occurring in the early
Universe. The cosmological constant ($\Lambda$) is found to decay
with the Hubble constant ($H$) as, $\Lambda\propto H^4$. In this
scenario the cosmological constant varies quadratically with the
energy density ($\rho$), i.e., $\Lambda\propto \rho^2$. Such a
variation is found to describe a two-component cosmic fluid in the
Universe. One of the component accelerated the Universe in the early
era, and the other in the present era. The scale factor of the
Universe varies as $a\sim t^{n}$, $n=\frac{1}{2}$ in the radiation
era. The cosmological constant vanishes when $n=\frac{4}{3}$ and
$n=\frac{1}{2}$. We have found that the inclusion of the term $R^2$
mimics a cosmic matter that could substitute the ordinary matter.
\end{abstract}
 \maketitle
\section{Introduction}
Modified gravity models have been invoked to resolve cosmological
and astrophysical problems with observations (Hawking, Luttrel,
Whitt, Srivastava, Sinha, Kung). The generalized Einstein action
including and additional scalar term $R^2$ is given by (Kenmoku et
al., 1992) by
\begin{equation}
S=-\frac{1}{16\pi G}\int d^4x\sqrt{g}(R+2\Lambda+\beta R^2\ )
\end{equation}
where $R$: Ricci scalar curvature, $\Lambda$: the cosmological
constant, $g$ : the negative determinant of the metric tensor
$g_{\mu\nu}$ and $\beta$ is a constant. Several authors have studied
classical solutions of this action without matter and have concluded
that big bang singularity may be avoided (see Kung, 1996). In this
paper we will study the cosmological implications of this action.

The variation of the metric with respect to $g_{\mu\nu}$ gives
\begin{equation}
R_{\mu\nu}-\frac{1}{2}Rg_{\mu\nu}-\Lambda g_{\mu\nu}+\beta
B_{\mu\nu}=-8\pi G T_{\mu\nu}\ ,
\end{equation}
where $T_{\mu\nu}$ is the energy momentum tensor of the cosmic
fluid, and
\begin{equation}
B_{\mu\nu}=2R(R_{\mu\nu}-\frac{1}{2}Rg_{\mu\nu})+2(R_{;\
\mu\nu}-g_{\mu\nu}\Delta R),
\end{equation}
with $R_{;\ \mu\nu}=\nabla_\mu\nabla_\nu R$ and $\Delta
R=g^{\mu\nu}R_{;\ \mu\nu}$. For an ideal fluid one has
\begin{equation}
T_{\mu\nu}=(\rho+p)u_\mu u_\nu+pg_{\mu\nu},
\end{equation}
where $u_\mu, \rho, p$ are the velocity, density and pressure of
the cosmic fluid. The flat Robertson-Walker line element is given
by
\begin{equation}
ds^2=-dt^2+a^2(t)\left(dr^2+r^2(d\theta^2+r^2\sin^2\theta\
d\phi^2)\right)
\end{equation}
The time-time and space-space components of Eq.(2) give
\begin{equation}
3H^2-\Lambda-18\beta\left(6\dot HH^2+2H\ddot H-\dot
H^2\right)=8\pi G\rho,
\end{equation}
and
\begin{equation}
-2\dot H-3H^2+\Lambda+6\beta\left(2\dddot H+12H\ddot H+18\dot
HH^2+9\dot H^2\right)=8\pi Gp,
\end{equation}
where $H=\frac{\dot a}{a}$ is the Hubble constant.
\section{Model A}
Now consider the cosmological model when
\begin{equation}
\Lambda=-18\beta\left(\dot H H^2+2H\ddot H-\dot H^2\right)
\end{equation}
so that Eq.(6)
\begin{equation}
3H^2=8\pi G\rho.
\end{equation}
and Eq.(7) becomes
\begin{equation}
-2\dot H-3H^2=8\pi G \left[p-\frac{3\beta}{2\pi G}\left(\dddot
H+3H\ddot H+12\dot H^2\right)\right],
\end{equation}
A universe with bulk viscosity ($\eta$) is obtained  by  replacing
the pressure $p$ by  the effective pressure  $p-3\eta H$. In this
case, one may attribute that the inclusion of the $R^2$ is
equivalent to having a bulk viscosity given by
\begin{equation}
\eta=\frac{\beta}{2\pi G}\left(\frac{\dddot H}{H}+3\ddot
H+12\frac{\dot H^2}{H}\right).
\end{equation}
However for a power law expansion of the form
\begin{equation}
a=A\ t^n, \qquad A\ , n \rm  \ \ \ const.
\end{equation}
one has $\dddot H+3\ddot HH=0$, so that
\begin{equation}
\eta=\left(\frac{6\beta\ n}{\pi G}\right)t^{-3}.
\end{equation}
The cosmological constant then becomes
\begin{equation}
\Lambda=\left(\frac{54\beta}{ n^3}\right)t^{-4}, \qquad n\ne 0,
\end{equation}
and the energy density
\begin{equation}
8\pi G\rho=\frac{3n^2}{t^2}.
\end{equation}
Using Eq.(12) the cosmological constant becomes
\begin{equation}
\Lambda=\left(\frac{54\beta}{n}\right)H^4, \qquad n\ne 0.
\end{equation}
Upon using Eq.(9), this becomes
\begin{equation}
\Lambda=\frac{6\beta}{n^6}\left(8\pi G\right)^2\rho^2, \qquad n\ne
0.
\end{equation}
Substituting Eq.(12) in Eq.(10), we see that the pressure is given
by
\begin{equation}
8\pi p=\left(\frac{(2-3n)n}{t^2}+\frac{72\beta\ n
(1-2n)}{t^4}\right).
\end{equation}
Using Eq.(15), this can be written as
\begin{equation}
p=\left(\frac{2}{3n}-1\right)\rho-\left(\frac{1-2n}{n^3}\right)N\
\rho^2, \qquad N=64\pi G\beta\qquad n\ne 0.
\end{equation}
We know the van der Waals equation of state is given by
\begin{equation}
p=\frac{\gamma\rho}{1-b\gamma}-\alpha \rho^2, \qquad \gamma,\ b,\
\alpha=\rm\ const.
\end{equation}
Thus, the resulting equation of state of a a power law expansion
is that due to two-component fluid resembling the van der Waals
equation of state. Therefore, introducing a term of $R^2$ in the
Einstein action is like introducing two fluid components in the
Universe. We see that one component of the fluid drives the
Universe in cosmic acceleration, by making $p<0$, in some period
and decelerates it in another period ($p>0)$. In the early
Universe, when the density was so huge, $p$ was negative if
$n<\frac{1}{2}$. During the matter dominated epoch, when the
density is very small, $p<0$, if \ $n>\frac{2}{3}$. Hence, we see
that the Universe accelerates for any deviation from the
Einstein-de Sitter expansion.
\subsection{Inflationary Era}
We see from Eq.(9) when $H=H_0=\rm\ const.$, i.e., $a\propto
\exp(H_0t)$, the cosmological constant vanishes, i.e.,
$\Lambda=0$. From Eq.(11) the bulk viscosity also vanishes, i.e.,
$\eta=0$. We recover the de Sitter solution, $p=-\rho$ [Eqs.(9)
and (10)]. For static universe $n=0$, the cosmological constant,
the bulk viscosity, the energy density and the pressure vanish,
i.e., $\Lambda=\eta=p=\rho=0$.
\subsection{Radiation Dominated Era}
During the radiation dominated phase, as in the Einstein-de Sitter
model, i.e., $a\propto t^{\frac{1}{2}}$, one has $n=\frac{1}{2}$ so
that Eq.(19) give the equation of the state $p=\frac{1}{3}\rho$.
Thus  the Einstein-de Sitter model is recovered. In this epoch the
cosmological constant vanishes. However, Eq.(19) shows that any
deviation form $n=\frac{1}{2}$ in the radiation era, viz.,
$n<\frac{1}{2}$, the second term will be large and negative. Thus,
an accelerated expansion of the Universe will be inevitable.
\subsection{Matter Dominated Era}
In the matter dominated epoch of Einstein-se Sitter model one has
$n=\frac{2}{3}$. In this case $p=72\pi G\beta\ \rho^2$. Since $\rho$
is small today, we see that the Universe asymptotically reduces to
Einstein-de Sitter type. However, for any deviation of the this
expansion law, $n>\frac{2}{3}$ accelerated expansion will be
inevitable. In this case $p<0$. So, in the distant future, when
$\rho\rightarrow 0$, the equation of  state reduces to
\begin{equation}
p=(-1+\frac{2}{3n})\ \rho=\omega \ \rho\ \qquad
(\omega=-1+\frac{2}{3n}).
\end{equation}
Thus, $n>\frac{2}{3}$ implies $\omega >-1$. We remark here in the
distant future, when $n\rightarrow\infty$, $p=-\rho$. Hence, the
future of our Universe will be a de-Sitter expansion.
\section{ Model B}
Now, let us define the cosmological constant by
\begin{equation}
\Lambda=-6\beta\left(2\dddot H+12H\ddot H+18\dot HH^2+9\dot
H^2\right),
\end{equation}
so that Eq.(6) and (7) become
\begin{equation}
3H^2=8\pi G(\rho+\bar\rho),
\end{equation}
and
\begin{equation}
-2\dot H-3H^2=8\pi Gp,
\end{equation}
where
\begin{equation}
8\pi G\bar\rho=-12\beta(\dddot H+3H\ddot H+6\dot H^2)
\end{equation}
\subsection{Inflationary Era}
We see that when $H=H_0=\rm \ const.$, i.e., $a\propto
\exp(H_0t)$, $p=-\rho$, $\Lambda=0$ and $\bar\rho=0$.
\subsection{Radiation and Matter dominated Eras}
Now consider a power law expansion of the scale factor of the form
as in Eq.(12). We find
\begin{equation}
8\pi Gp=n(2-3n)\ t^{-2},
\end{equation}
\begin{equation}
\Lambda=18\beta \ n (2n-1)(3n-4)\ t^{-4},
\end{equation}
and
\begin{equation}
\rho= \left(\frac{3n}{2-3n}\right)
p+N'\left(\frac{2n-1}{n^2(2-3n)^2}\right)p^2,\ N'=637 \ \beta \ G
, \ n\ne \frac{2}{3}
\end{equation}
Eq.(28) represent our equation of state for the present cosmology.
The cosmological constant here varies as
\begin{equation}
\Lambda=\frac{243}{2}\beta \ H^4,
\end{equation}
The equation of state now reads,
\begin{equation}
p=\omega(t)\rho, \qquad \omega(t)=\left(\frac{3n}{2-3n}+72\
\beta\frac{(2n-1)}{n(2-3n)}\frac{1}{t^2}\right)^{-1}.
\end{equation}
It is evident when $n\rightarrow\infty $ ( i.e.,
$a\rightarrow\infty),\ \omega\rightarrow -1$ and the Universe
becomes vacuum dominated an expands like  de-Sitter. It is
interesting to note that when $n=\frac{4}{3}$, the cosmological
constant vanishes, i.e., $\Lambda=0$. In this case the pressure
becomes negative, i.e., $p<0$, and  this drives the Universe into an
epoch of cosmic acceleration. The energy density becomes
\begin{equation}
\rho=c_1H^2+c_2H^4\ , \qquad c_1\ ,\ c_2\rm \ \ consts.
\end{equation}
The deceleration parameter $q=-\frac{\ddot a}{a\ H^2}=-0.25$. Once
again, when $n=\frac{1}{2}$, we recover the Einstein-de Sitter
solution, i.e., $a\propto t^{\frac{1}{2}}$ and $p=\frac{1}{3}\
\rho$ and $\Lambda=0$. For $n=\frac{2}{3}$, $p=0$,
$\Lambda=-8\beta t^{-4}$, and $8\pi G\rho=\frac{4}{3
t^2}\left(1+\frac{12\beta}{t^2}\right)$. Hence, the Universe
approaches the Einstein-de Sitter solution asymptotically
($t\rightarrow\infty$).
\section{Model C}
Now consider a cosmological model in which $\Lambda=0$. In this
case, eqs.(6) and (7) yield
\begin{equation}
3H^2=8\pi\ G(\rho+\rho')
\end{equation}
and
\begin{equation}
-2\dot H-3H^2=8\pi G(p+p'),
\end{equation}
where 
\begin{align}
8\pi G\rho'&=18\beta (6\dot HH^2+2H\ddot H-\dot H^2), \\
\nonumber 8\pi Gp'&=-6\beta (2\dddot H+12H\ddot H+18\dot
HH^2+9\dot H^2),
\end{align}
Eqs.(32) and (33) can be written as
\begin{equation}
3H^2=8\pi\ G\rho_{\rm eff.}
\end{equation}
and
\begin{equation}
-2\dot H-3H^2=8\pi Gp_{\rm eff.},
\end{equation}
where
\begin{equation}
\rho_{\rm eff.}=\rho+\rho'\ , \qquad p_{\rm eff.}=p+p'\ .
\end{equation}
We, therefore, argue  that the inclusion of the term $R^2$ in the
Einstein action induces a fluid in the Universe that has pressure
$(p')$ and energy density ($\rho'$), in addition to the preexisting
matter. This may suggest that our Universe is  fluid with two
components; one is \emph{bright} ($\rho$) and the other is
\emph{dark} ($\rho')$ without having a cosmological constant.
\subsection{The Inflationary Era}
An inflation solution arises when $H=H_0=\rm\ const.$ which is
solved to give $a\propto \exp(H_0t)$. Eq.(35) and (36) give
\begin{equation}
p_{\rm eff.}=-\rho_{\rm eff.}
\end{equation}
With some scrutiny, one would discover that Eq.(34) implies that
$p'=\rho'=0$. Hence, the \emph{dark} component does not contribute
to this inflationary era.
\subsection{Radiation Dominated Era}
Consider now a power law expansion for the Universe during the
radiation dominated era of the form
\begin{equation}
a=B\ t^n\ , \qquad n\ , B,\ \ {\rm \ \ const.}
\end{equation}
Substituting this in Eq.(34) one gets
\begin{equation}
\rho'=\frac{54\beta\ n^2(1-2n)}{8\pi G}t^{-4},
\end{equation}
and
\begin{equation}
p'=\frac{18\beta\ n(1-2n)(4-3n)}{8\pi G}t^{-4}.
\end{equation}
The two equations are related by the relation
\begin{equation}
p'=\left(-1+\frac{4}{3n}\right) \ \rho'\ , \qquad n\ne 0\ ,
\end{equation}
which represents the equation of stat of the \emph{dark} fluid. It
is very interesting to note that, when $n=\frac{1}{2}$,
$\rho'=p'=0$. This implies that the \emph{dark} component does not
disturb the nucleosynthesis constraints set forth bye the
Einstein-de Sitter solution. Notice that when $n\rightarrow\infty$,
$p'=-\rho'$, so that in the distant future the our Universe with or
without normal matter will be vacuum dominated. We notice that this
\emph{dark} component does not live in a static Universe since it
has $p'=\rho'=0$. For a positive energy density, we have (for
$\beta>0)$ the constraint $n< \frac{1}{2}$. For $\frac{1}{2}
<n<\frac{4}{3}$, $\rho'<0 , \ p' <0$. The positivity of energy
density is recovered if $\beta<0$. It is remarkable to notice that
if the matter action is not incorporated in the Einstein action, the
inclusion of the quadratic term acts like matter. This type of
matter is characterized by its equation of state in Eq.(42). Hence,
the inclusion of the $R^2$ mimics the introduction of a new matter
in the Universe.
\subsection{Matter Dominated Era}
Applying Eqs.(39)-(41) into Eqs.(32) and (33), one gets
\begin{equation}
8\pi
G\rho=\frac{3n^2}{t^2}\left(1-\frac{18\beta(1-2n)}{t^2}\right),
\end{equation}
and
\begin{equation}
8\pi
Gp=\frac{n(2-3n)}{t^2}\left(1-\frac{18\beta(2n-1)(3n-4)}{(2-3n)}\frac{1}{t^2}\right),
\end{equation}
Once again, we see  from Eq.(41) that the pressure of the
\emph{dark} fluid vanishes when $n=\frac{4}{3}$ leaving only the
\emph{bright} fluid to contribute to the universal pressure. In this
case, $p<0$, and this will drive the Universe into a cosmic
acceleration era. Substitution of  $n=\frac{2}{3}$ in Eqs.(43) and
E(44) yields
\begin{equation}
8\pi G\rho=\frac{4}{3t^2}\left(1+\frac{6\beta}{t^2}\right),\qquad
8\pi G p=\frac{8\beta}{t^4}\ .
\end{equation}
We remark that the Universe approaches the Einstein-de Sitter
asymptotically (when $t\rightarrow\infty$, i.e., in the distant
future), where $\rho=\frac{1}{6\pi Gt^2}$ and $p=0$. However,
$\rho_{\rm eff.}=\frac{1}{6\pi Gt^2}$ and $p_{\rm eff.}=0$. Notice
that during this era the \emph{dark} component behaves like a
stiff matter, with $p'=\rho'$. It is evident that when
$\frac{2}{3} <n<\frac{4}{3}$ the Universe enters a period of
cosmic acceleration. We may thus argue that the present observed
cosmic acceleration happened during this period. Eq.(45) looks
like having a cosmological constant of the form $\Lambda\propto
H^4$ in the Einstein-de Sitter universe.
\section{Acknowledgements}
I gratefully acknowledge the financial support of the Swedish
International  Development Cooperation Agency (SIDA), and the the
Abdus Salam International Center for Theoretical Physics for
hospitality, where this work is carried out during my visit to  the
center as  a Regular Associate.
\section{References}
\hspace{-0.38cm}Hawking, S.W., and Luttrel, J.C., \emph{Nucl. Phys.
B247}, 250 (1984).\\
Kenmoku, M., Kitajima, E., and Okamoto, Y., \emph{hep-th}/9203226 (1992).\\
Kung, J.H., \emph{Phys. Rev. D53}, 3017 (1996).\\
Nojiri, N. and Odintsov, S.D., \emph{Phys. Red. D72}, 023003 (2005).\\
Robert, D.M., \emph{Mon. Not. Roy. Soc. Lond. 249}, 339 (1991).\\
Srivastava, S.K., \emph{Phys. Lett. B643}, 1 (2006): \emph{Phys. Lett. B648}, 119 (2007).\\
Srivastava, S.K., and Pinha, K.P., \emph{Phys. Lett. B307}, 40 (1993).\\
Whitt, B.S.,  \emph{Phys. Lett. 145B}, 176 (1984).\\
\end{document}